\documentclass{jpsj-suppl}
\usepackage{graphicx}
\usepackage{epsf}
\def\lamb#1#2{$^{#1}_{\Lambda}${#2}}

\def\la#1#2{$^{#1}_{~~\Lambda}${#2}}

\title{Tribute to Tullio Bressani, Bogdan Povh and Toshimitsu Yamazaki 
} 
\author{Avraham \textsc{Gal}} 
\inst{Racah Institute of Physics, The Hebrew University, Jerusalem 91904, 
ISRAEL} 

\recdate{January 19, 2026} 

\abst{In this HYP25 talk I pay tribute to Tullio Bressani (1940-2024), Bogdan 
Povh (1932-2024) and Toshimitsu Yamazaki (1934-2025), all of whom made lasting 
contributions to shaping up Strangeness Nuclear Physics. Yoshinori Akaishi's 
(1941-2025) record is also noted.} 

\kword{strangeness nuclear physics} 

\begin{document}
\maketitle

\section{Introduction} 

\begin{figure}[!ht] 
\begin{center} 
\includegraphics[scale=0.4]{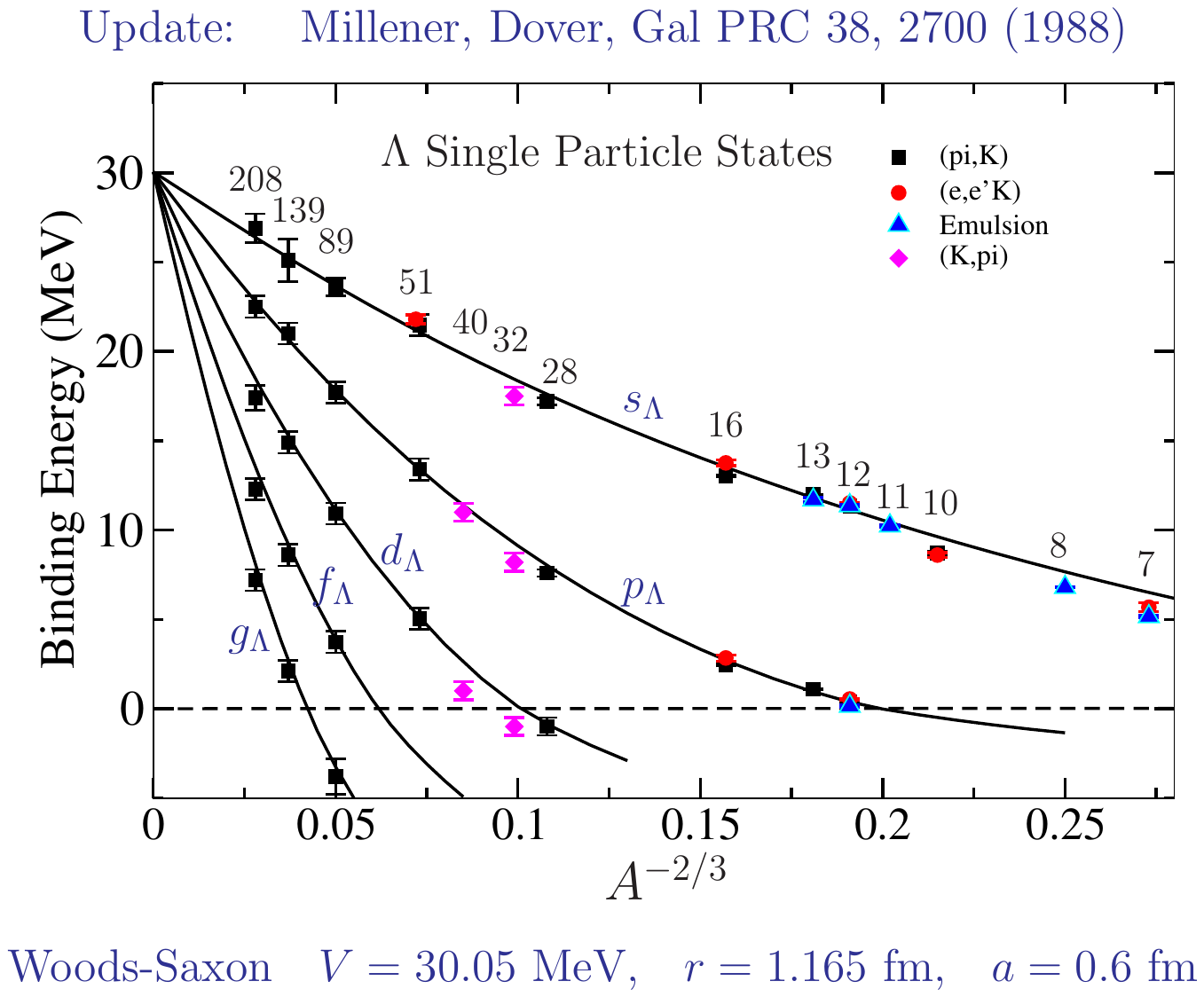} 
\caption{$B_\Lambda$ values in \lamb{7}{Li} to \la{208}{Pb} from experiment, 
and as calculated from a 3pF WS potential, suggesting a $\Lambda$-nucleus 
potential depth $D_{\Lambda}\approx -30$~MeV. Figure adapted from 
Ref.~\cite{GHM16}.} 
\label{fig:MDG} 
\end{center} 
\end{figure}

The modern period in hypernuclear physics started in the early 1970s using the 
in-flight $(K^-,\pi^-)$ production reaction at CERN by Bressani's and Povh's 
groups and later on by Chrien's group at BNL. It was followed in 1985 by the 
$(\pi^+,K^+)$ production reaction at BNL and applied since 1995 at KEK, led 
by Yamazaki. The resulting $\Lambda$-hypernuclear g.s. binding energies form 
large fraction of the world data shown in Fig.~\ref{fig:MDG}. 
Below I summarize major physics contributions made by Bressani, Povh and 
Yamazaki, noting also Akaishi's record. 

\section{The early years} 
\label{sec:early} 

Bressani and Povh, separately, started $(K^-,\pi^-)$ counter experiments at 
the CERN-PS as documented for $K^-$ in-flight in the following publications.  
\begin{itemize} 
\item 
\textbf{Bressani's group}: G.C.~Bonazzola, et al. ($p_{K^-}$=390~MeV/c), 
\textit{Production of \lamb{12}{C} by $K^-$ in flight}, PLB \textbf{53} (1974) 
297; \textit{Observation of \lamb{16}{O} and \lamb{27}{Al}}, PRL \textbf{34} 
(1975) 683. However, their energy resolution of about 6 MeV was insufficient 
to draw strong conclusions. 
\item 
\textbf{Povh's group}: W.~Br\"{u}ckner, et al. ($p_{K^-}$=900~MeV/c), 
\textit{Hypercharge exchange reaction on nuclei}, PLB \textbf{55} (1975) 107; 
\textit{Strangeness exchange reaction on nuclei}, PLB \textbf{62} (1976) 481; 
followed by the Heidelberg-Saclay-Strasbourg collaborative work at 
$p_{K^-}$=715 MeV/c, reaching 2 MeV energy resolution: \textit{Spin-orbit 
interaction of $\Lambda$ particles in nuclei}, PLB \textbf{79} (1978) 157; 
and many more Bertini, et al., papers in the 1980s~\cite{GHM16}.  
\end{itemize} 
Both Povh at CERN and Yamazaki at the Tokyo-INS and in KEK were looking for 
$\Sigma$ hypernuclei in $(K^-/K^-_{\rm stop},\pi^-)$ reactions during the 
1980s, but the statistical significance of the signals observed was rather 
poor. It took some time to realize that the $\Sigma$-nuclear interaction is 
largely repulsive so $\Sigma$ hypernuclei generally do not exist~\cite{DMG89} 
except for ${_{\Sigma}^4}$He established by Hayano, et al.~\cite{Hayano89}, 
in a $^4{\rm He}(K^-_{\rm stop},\pi^-)$ reaction at KEK and soon explained 
in a work led by Akaishi~\cite{Harada90}. Povh subsequently left hypernuclear 
physics in favor of other topics, notably the EMC effect, whereas Yamazaki 
became fully involved in hypernuclear physics around 1985. As for Bressani, 
he moved on to other topics at CERN, notably antineutrons, returning to 
hypernuclear physics around 2000 during which he was leading FINUDA at 
DA$\Phi$NE, Frascati, using their high-quality $\phi$-factory $K^-_{\rm stop}$ 
source.

\section{Tullio Bressani (1940-2024)}
\label{sec:tullio} 

\begin{figure}[!h] 
\begin{center} 
\includegraphics[scale=0.6]{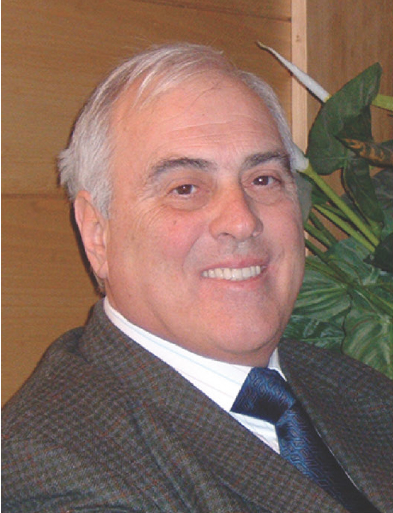}
\caption{Tullio Bressani (1940-2024)} 
\label{fig:TB} 
\end{center} 
\end{figure} 

\subsection{CV highlights} 
\label{subsec:TBcv} 
\begin{itemize} 
\item 
M. Sc. -- Universit{\'a} degli Studi di Torino 1962. 
\item 
Assoc. Prof. -- Cagliari 1964-1968, Torino 1968-1976. 
\item 
Professor -- Cagliari 1976-1984, Torino 1984-2010. 
\item 
Chair: HADRON93 (Como); HYP88/00 (Padova/Torino); Winter School Hadron Physics 
(Folgaria: 1985-1993); Enrico Fermi International School (Varenna: Hadron 
Physics 2004, Strangeness-Spin Physics 2007). 
\item 
Head of the OBELIX, FENICE and PAN facilities at CERN (1980s \& 1990s) as 
reviewed by T. Bressani and A. Filippi, \textit{Antineutron Physics}, 
Phys. Rep. \textbf{383} (2003) 213-297. Head of the FINUDA collaboration 
formed around the end of 1990s at DA$\Phi$NE, Frascati. 
\item 
Member of the 1st J-PARC PAC (early 2000s) and of many more Scientific 
Committees. 
\end{itemize}

\subsection{Major FINUDA publications} 
\label{subsec:FINUDA} 
\begin{itemize} 
\item 
M. Agnello, et al. (FINUDA + A. Gal), \textit{New results on MWD of p-shell 
$\Lambda$ hypernuclei}, PLB \textbf{681} (2009) 139.
\item 
M. Agnello, et al. (FINUDA), \textit{Hypernuclear spectroscopy with 
$K^-_{\rm stop}$ on $^7$Li, $^9$Be, $^{13}$C, $^{16}$O}, PLB \textbf{698} 
(2011) 219. 
\item 
M. Agnello, et al. (FINUDA + A. Gal), \textit{Evidence for heavy 
hyperhydrogen \lamb{6}{H}}, PRL \textbf{108} (2012) 042501.
\item 
M. Agnello, et al. (FINUDA), \textit{First determination of the proton induced 
NMWD width of p-shell $\Lambda$ hypernuclei}, PLB \textbf{738} (2014) 499. 
\item 
E. Botta, T. Bressani, and A. Feliciello (FINUDA), \textit{On binding energy 
and CSB in $A \leq 16$ $\Lambda$ hypernuclei}, NPA \textbf{960} (2017) 165. 
\end{itemize}

\section{Bogdan Povh (1932-2024)}
\label{sec:bogdan} 

\begin{figure}[!h] 
\begin{center} 
\includegraphics[scale=0.5]{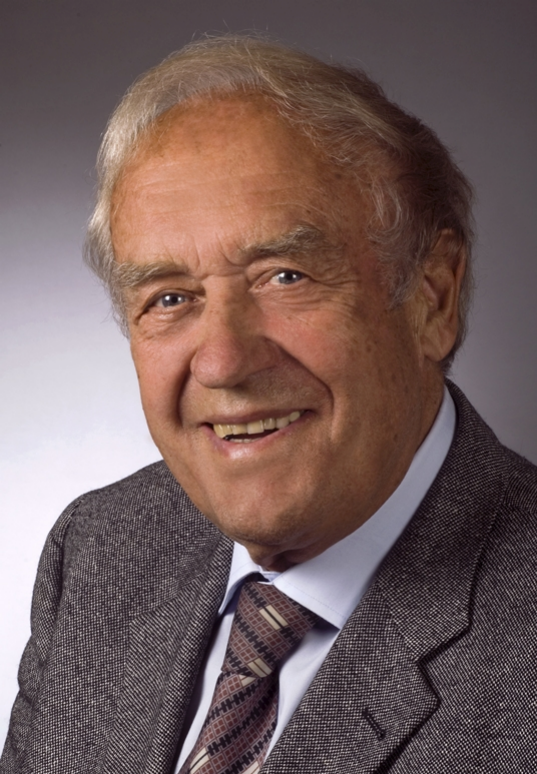}
\caption{Bogdan Povh (1932-2024)} 
\label{fig:BP} 
\end{center} 
\end{figure} 

\subsection{CV highlights} 
\label{subsec:BPcv} 
\begin{itemize} 
\item 
1955 -- Diploma in Physics, Univ. Ljubljana. 
\item 
1957-1959 -- Research Fellow, CALTECH (Pasadena). 
\item 
1960-1962 -- Assistant, Univ. Ljubljana (1960: Ph.D.). 
\item 
1962-1964 -- Assistant, Univ. Freiburg (1964: Habilitation). 
\item 
1965 -- Associate Professor, Univ. Heidelberg. 
1966-2000 -- Professor, Univ. Heidelberg. 
\item 
Since 1975: Director, MPI f. Kernphysik, Heidelberg. 
\item 
Chair: First HYP conference, HYP82 at Heidelberg. 
\item 
Sabbaticals: CERN, LBL, LANL, UC-Berkeley. 
\item 
Editor: 1978-2000 Zeit. Phys. A (EPJA since 1998). 
\item 
Author: several widely used High-Energy Nucl. Phys. textbooks. 
\item 
Since 2000: Prof. Emeritus; 2005: Stern-Gerlach Medal. 
\end{itemize} 
Povh's $(K^-,\pi^-)$ experiments at the CERN-PS in the 1970s, and together 
with Bertini (Saclay) in the 1980s, established hypernuclear spectroscopy by 
observing systematically bound-state and continuum spectra of hypernuclei from 
\lamb{9}{Be} to \lamb{40}{Ca}.

\section{Toshimitsu Yamazaki (1934-2025)} 
\label{sec:toshi} 

\begin{figure}[!h] 
\begin{center} 
\includegraphics[scale=2.0]{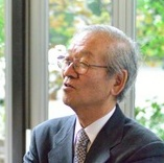}
\caption{Toshimitsu Yamazaki (1934-2025)} 
\label{fig:TY} 
\end{center} 
\end{figure} 

Toshimitsu Yamazaki was leading Strangeness Nuclear Physics worldwide,
Japan in particular, for several decades. It is reflected today in the rich
\& imaginative experimental program in J-PARC, where many of his students
have become leaders on their own right. 

Yamazaki's leading contributions to Hadronic Physics, in particular to
Strangeness Nuclear Physics, are classified in the Japan Academy website by 
five Recurring Themes. 
\begin{itemize}
\item I.~~~Meson effects in nuclear magnetic moments.
\item II.~~Muon spin rotation ($\mu$SR) relaxation.
\item III.~Discovery of metastable antiprotonic helium.
\item IV.~Discovery of deeply bound pionic-atom states.
\item V.~~Search for kaonic nuclei; Kaonic Proton Matter (KPM).
\end{itemize}
Here I focus on the last two Recurring Themes, developed and matured
throughout the last 40 years since I first met him at TRIUMF in 1985
on my sabbatical stay there.

\subsection{Deeply bound pionic atoms}
\label{subsec:pionic} 

\begin{figure}[hbt] 
\begin{center} 
\includegraphics[scale=0.40,clip]{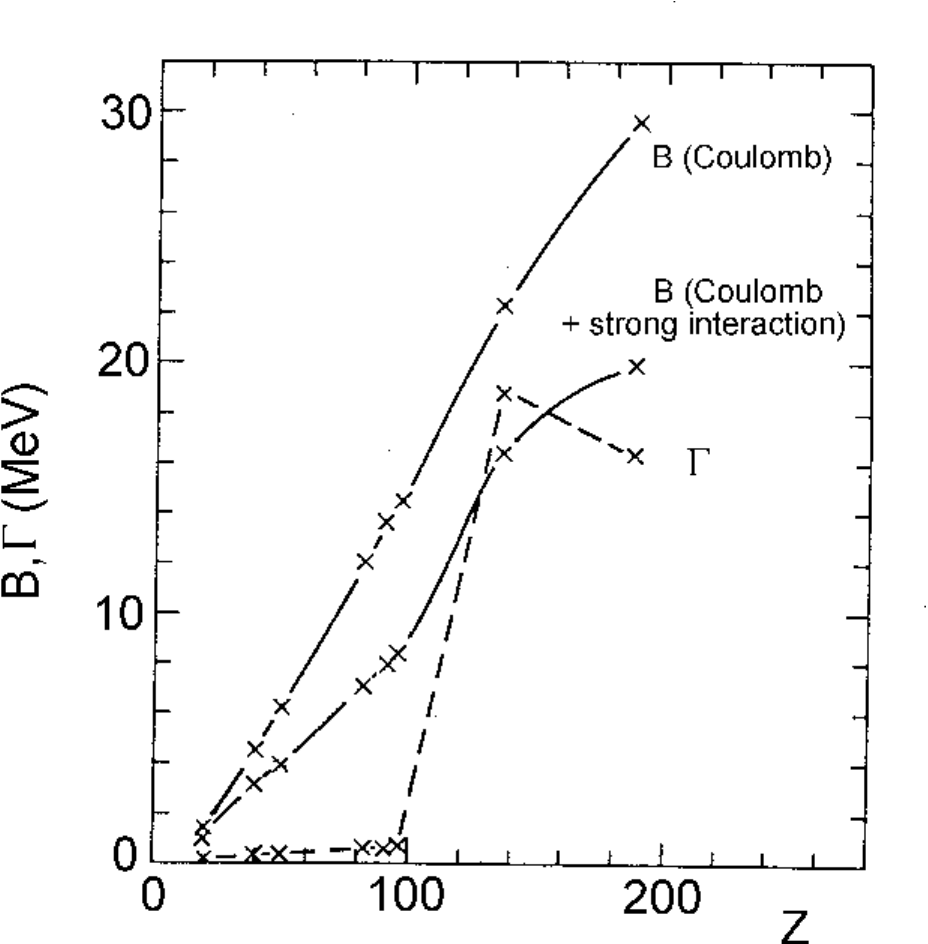}
\includegraphics[scale=0.60,height=7cm,clip]{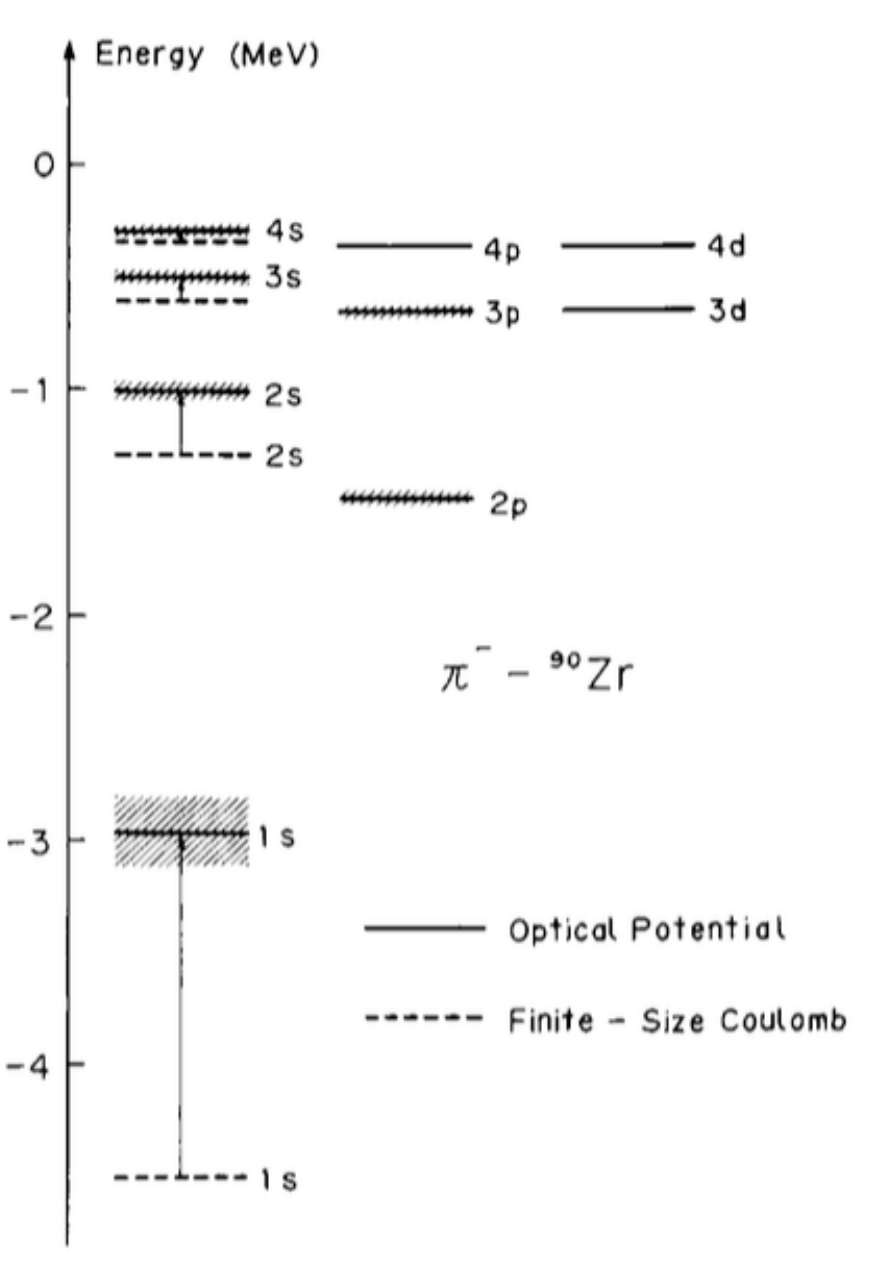}
\caption{Left: binding energies (B) and widths ($\Gamma$) of 1s states in
pionic atoms (Friedman \& Soff~\cite{FS85}). Right: $\pi^-$ energy levels in
$^{90}$Zr with \& without Strong Interactions (Figure adapted from Toki \&
Yamazaki~\cite{TY88}).}
\label{fig:1s}
\end{center}
\end{figure}

Deeply bound pionic atoms refer to 1s and 2p states in heavy pionic atoms
which cannot be reached in X-ray cascade because upper levels such as 3d
are already broadened by Strong Interactions rendering the radiative yield
exceedingly small. In 1985, Friedman and Soff~\cite{FS85} noted that deeply
bound 1s pionic states are sufficiently narrow to make them well defined,
see Fig.~\ref{fig:1s} (left) where the calculated 1s widths remain small
up to the top end of the periodic table. The relative narrowness of the 1s
states follows from the {\it repulsive} real part of the s-wave pion-nucleus
potential that keeps the absorptive imaginary part off the nuclear volume. 

Three years later in 1988 Toki and Yamazaki~\cite{TY88}, unaware of the
Friedman-Soff paper, realized that the width of 1s pionic states in heavy
nuclei is considerably smaller than the 1s$\to$2p excitation energy, see
Fig.~\ref{fig:1s} (right), so it made sense to look for strong-interaction
reactions to populate such states. It took another eight years to apply
a good `recoil-less' $^{208}$Pb(d,$^3$He) reaction at beam energy about 600
MeV in GSI to create a deeply bound pion as close to rest as possible in
$^{207}$Pb (Yamazaki, Hayano, et al.~\cite{YH96}). This was followed by
experiments on $^{206}$Pb and on Sn isotopes, as reviewed by Kienle and
Yamazaki~\cite{KY04}, Friedman and Gal~\cite{FG07} and Yamazaki, Hirenzaki,
Hayano and Toki \cite{YHHT12}. More recently, both 1s and 2p deeply-bound
pionic states in $^{121}$Sn were identified and studied at RIKEN~\cite{Sn18}.

Pionic-atom states, deeply-bound as well as `normal' ones observed in atomic
cascade, provide good evidence for a nuclear-medium renormalization of the
isovector s-wave $\pi N$ scattering length $b_1$ in a form suggested by
Weise~\cite{W00}:
\begin{equation}
b_1(\rho)=b_1^{\rm free}(1-{\frac{\sigma_{\pi N}}{m_{\pi}^2 f_{\pi}^2}}
\rho)^{-1},
\label{eq:W00}
\end{equation}
where $f_{\pi}=92.2$~MeV is the free-space pion decay constant and
$\sigma_{\pi N}$ is the pion-nucleon sigma term. Fitting globally pionic
atom energy shifts and widths across the periodic table, Friedman and Gal
were able to extract the value $\sigma_{\pi N}=57\pm 7$~MeV~\cite{FG19},
in excellent agreement with a value 59.1$\pm$3.5~MeV derived using $\pi^-$H
and $\pi^-$D atom data~\cite{Hofe15}, or 58$\pm$5~MeV derived using $\chi$EFT
$\pi N$ scattering lengths~\cite{Hofe18}.

\subsection{Kaonic Proton Matter}
\label{subsec:KPM}

Interpreting the $\Lambda$(1405) as a $K^-p$ quasibound state suggests that
$K^-$ mesons are likely to bind strongly into nuclear clusters, the simplest
of which is $K^-pp$. That a $J^{\pi}=0^-,I=\frac{1}{2}$ $K^-pp$ state might be
bound by $\sim$10~MeV was suggested by Nogami already in 1963~\cite{Nogami63}.
Yamazaki and Akaishi~\cite{YA02,AY02}, using a complex energy-independent
${\bar K}N$ potential within a single-channel $K^-pp$ calculation, obtained
binding energy $B_{K^-pp}\sim 50$~MeV and width $\Gamma_{K^-pp}\sim 60$~MeV.
Subsequent ${\bar K}NN$--$\pi\Sigma N$ coupled-channel Faddeev calculations
\cite{SGM07,SGMR07} confirmed this order of magnitude for $B$ but gave
larger width values ($\sim$100~MeV). Using a chirally motivated ${\bar K}N$
interaction in such Faddeev calculations lowers $B_{K^-pp}$ to about 32~MeV
and $\Gamma_{K^-pp}$ to about 50~MeV~\cite{RS14}, while few-body calculations
using a single-channel effective chirally motivated ${\bar K}N$ interaction
find even smaller values of binding energies and widths~\cite{DHW08,BGL12}.
Experimentally, following several dubious claims by several experiments,
J-PARC E15 reported a statistically reliable $K^-pp$ signal~\cite{Yam20} with
$B_{K^-pp}=42\pm 3^{+3}_{-4}$~MeV, $\Gamma_{K^-pp}=100\pm 7^{+19}_{-9}$~MeV.
Part of this large width comes from $K^- NN$ absorption processes that are
not accounted for in most calculations.

Calculations of $\bar K$ nuclear clusters up to six nucleons~\cite{Ohnishi17}
demonstrate that replacing one nucleon by a $\bar K$ meson increases
substantially the overall binding energy. This suggests to look for a maximal
increase by considering aggregates of bound $\Lambda^{\ast}$ hyperons,
so-called Kaonic Proton Matter (KPM) by Akaishi and Yamazaki~\cite{AY17} who
argued that it would also provide an absolutely stable form of matter for less
than ten $\Lambda^{\ast}$ hyperons. This exciting proposal was questioned by
the Jerusalem-Prague collaboration~\cite{Jarka18,Jarka19,Jarka20} within
a Relativistic Mean Field (RMF) calculational scheme, as follows.

\begin{figure}[htb] 
\begin{center} 
\includegraphics[scale=0.4,clip]{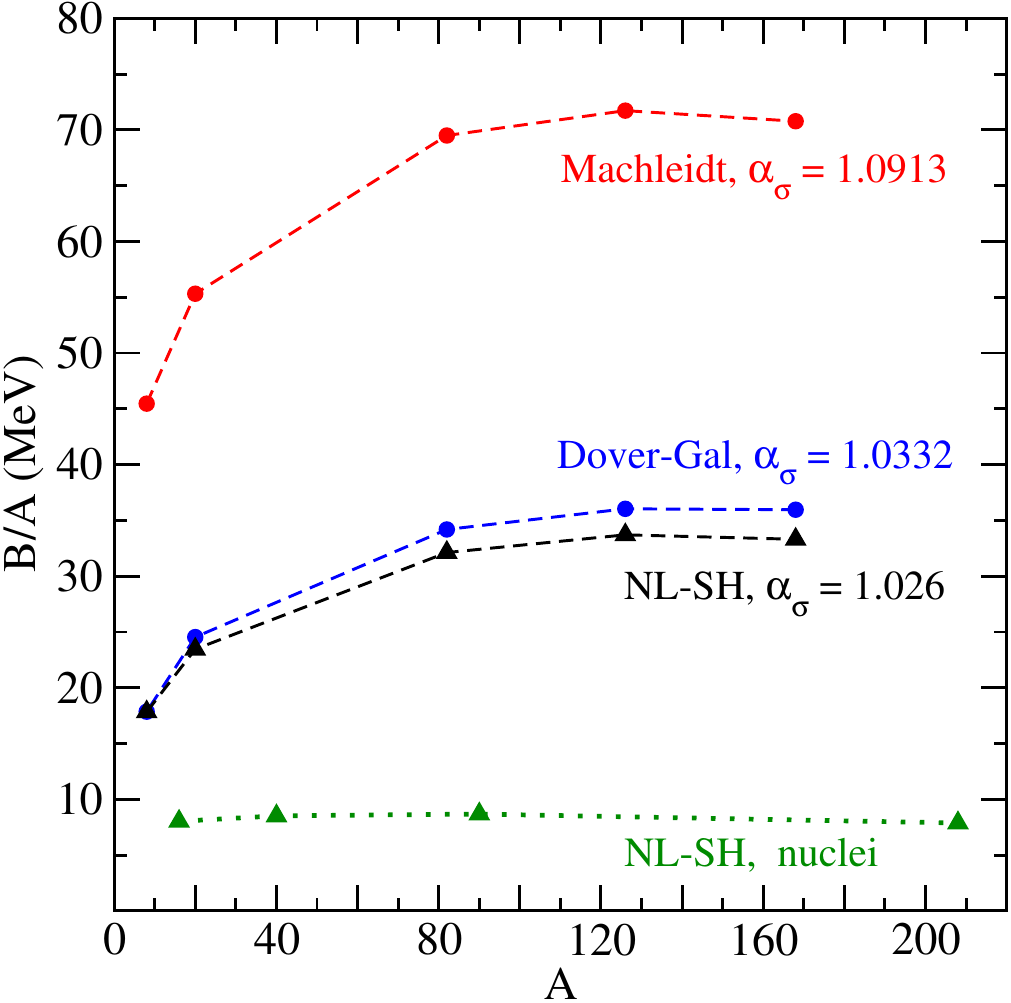} 
\includegraphics[scale=0.4,clip]{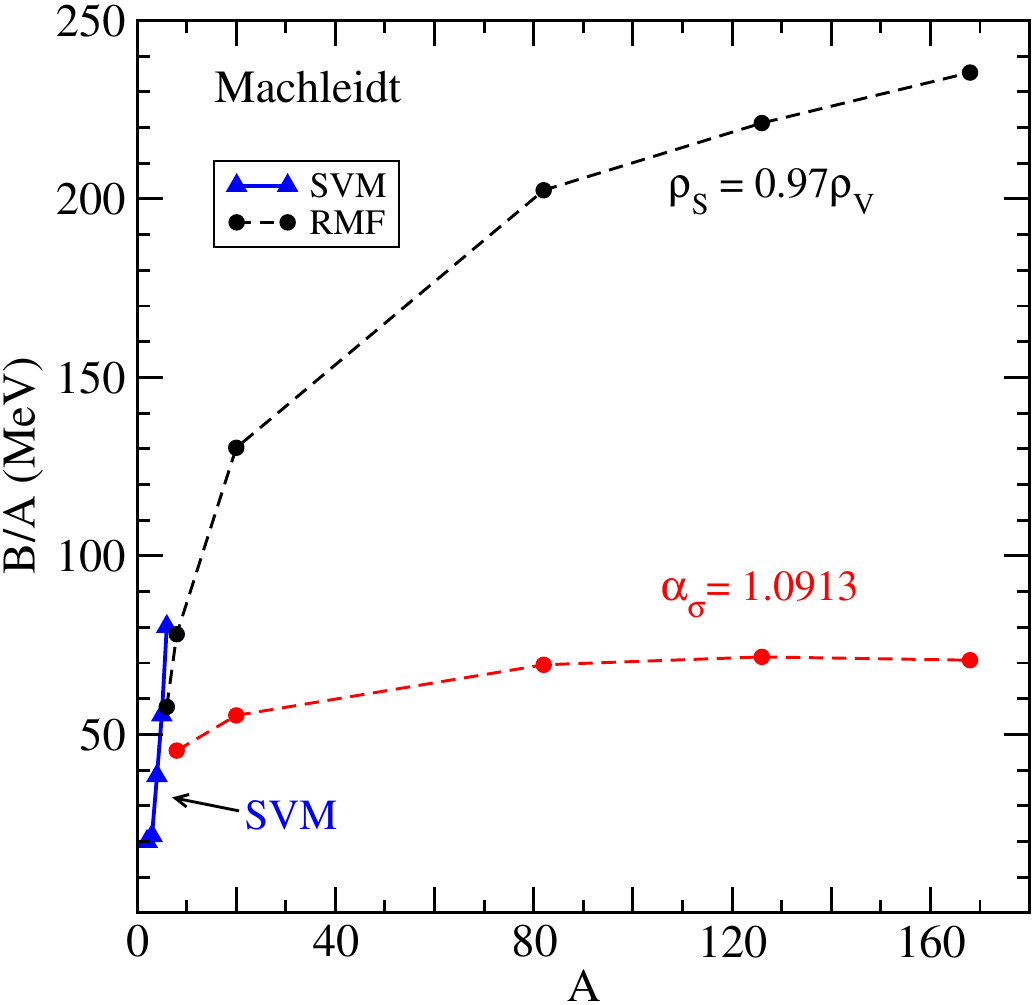} 
\caption{Left: binding energies per particle ($B/A$) of $\Lambda^{\ast}$ 
nuclei as function of mass number $A$, calculated in two RMF versions, 
saturating at $B/A \approx 70$ (Machleidt) or 35~MeV (Dover-Gal). $B/A$ values 
calculated for atomic nuclei are shown for comparison. Right: $B/A$ for 
$\Lambda^{\ast}$ nuclei from the left-hand side calculation (both in red), 
compared with a similar calculation (in black) using $\rho_s=0.97\rho_v$. 
$B/A$ values calculated in few-$\Lambda^{\ast}$ systems, marked SVM, 
are shown for comparison. Figure adapted from Fig.~2, Ref.~\cite{Jarka20}.} 
\label{fig:RMF} 
\end{center} 
\end{figure} 

The $\Lambda^{\ast}\Lambda^{\ast}$ interaction strength used as input to these
RMF calculations of $\Lambda^{\ast}$ nuclei was constrained by a two-body
binding energy value $B(\Lambda^{\ast}\Lambda^{\ast})\approx 40$~MeV, deduced
from $B(K^-K^-pp)=93$~MeV following an earlier work by Maeda, Akaishi and
Yamazaki~\cite{MAY13}. It was found then, as shown on the left-hand side of
Fig.~\ref{fig:RMF}, that the binding energy per $\Lambda^{\ast}$ saturates
at values well below 100 MeV for mass number $A\geq 120$, implying that
$\Lambda^{\ast}$ matter is highly unstable against strong-interaction decay
to $\Lambda$ and $\Sigma$ aggregates. Recall that about 300 MeV is required
to reduce the $\Lambda^{\ast}(1405)$ mass in the medium below that of the
lightest hyperon $\Lambda(1116)$. It is worth noting that RMF calculations for
multi $\bar K$-N-$\Lambda$ hadronic systems reported earlier by Gazda et
al.~\cite{GFGM07,GFGM08,GFGM09} reached similar conclusions, although in the
context of ruling out kaon condensation.

The saturation of $B/A$ observed on the left-hand side of Fig.~\ref{fig:RMF}
follows from the decrease of the scalar density $\rho_s$ associated with
the attractive $\sigma$ field with respect to the conserved vector density
$\rho_v$ associated with the repulsive $\omega$ field: 
$\rho_s=(M^{\ast}/E^{\ast})\rho_v < \rho_v$, 
where $M^{\ast}=M-g_{\sigma B}{\bar\sigma}$ is the baryon-$B$ effective mass. 
Thus, Lorentz invariance implies that the scalar-field attraction is reduced
as density increases. The upper curve on the right-hand side of
Fig.~\ref{fig:RMF} demonstrates then how a {\it fixed} ratio
$(\rho_s/\rho_v)=0.97$, corresponding to $^{16}$O calculation, leads to
non-saturating $B/A$ values in contrast to the saturated values obtained in
the properly calculated RMF lower curve. It is worth noting that the central
density of $\Lambda^{\ast}$ matter is found to saturate as well, at roughly
twice nuclear-matter density.

\section*{Appendix: Yoshinori Akaishi (1941-2025)} 

Yoshinori Akaishi passed away a short while before HYP25 was convened. 
I was asked by the Organizers to include him too in my Obituary talk, 
which I respectfully accepted to do. Akaishi-san made valuable contributions 
to Nuclear and Hadronic Physics in Japan and elsewhere. A brief CV follows. 
\begin{itemize} 
\item 
1969-1991: Hokkaido Univ, Sapporo (Ph.D. 1971 Kyoto). 
\item 
1992: Prof. INS, Univ Tokyo; 1997: INS $\to$ KEK. 
\item 
1999: retired; Nihon Univ; visitor at RIKEN. 
\item 
156 publications 1966-2024, mostly in Few-Body Physics. 
\item 
1981: 1st paper on $\Lambda$ hypernuclei (\lamb{5}{He}). 
\item 
1988: 1st paper on $\Sigma$ hypernuclei (${_{\Sigma}^9}{\rm Be}$). 
\item 
1997-2024: many papers with Yamazaki, mostly $\bar K$ nuclei. 
\end{itemize}

\section*{Acknowledgments}

I would like to thank the Organizers of HYP 2025, in particular Satoshi 
(Nue) Nakamura, for inviting me to give this Tribute talk. Special thanks are 
due to Alessandro Feliciello, Hans Weidenm\"{u}ller and Hiroyuki Noumi for 
providing me with biographies and recollections of the late Tullio Bressani, 
Bogdan Povh and Toshi Yamazaki, respectively. Thanks are due also to Eliahu 
Friedman, Jaroslava {\'O}bertov{\'a} (Hrt{\'a}nkov{\'a}) and Hiroshi Toki for 
providing me with adapted versions of published figures.

\end{document}